# Evaluation of waterway lock service quality in Yangtze Delta: from the perspectives of customer and supplier

Wenzhang Yang[1], Peng Liao[1*], Shangkun Jiang[2], Hao Wang[1]

[1] *School of Transportation, Southeast University, Nanjing, 211189, China*
[2] *Department of Civil and Coastal Engineering, University of Florida, Gainesville, FL, 32611, USA*

**ABSTRACT**

In recent decades, the waterway locks in the Yangtze Delta, China, have become major traffic bottlenecks. To gain a comprehensive understanding of the crew's perspectives and primary concerns regarding lock services during vessel lockage, and to enhance customer satisfaction and improve vessel lockage efficiency, it is necessary to assess the waterway lock service quality (WLSQ). This paper presents an evaluation system for WLSQ from various stakeholders' viewpoints. Firstly, by employing questionnaire surveys and the structural equation model method, in conjunction with factor analysis, the WLSQ and its influencing factors in the Yangtze River Delta region are analyzed from a customer perspective. Secondly, the Analytic Hierarchy Process method is utilized, along with a dedicated questionnaire for service suppliers, to examine their concerns regarding the performance of vessel lock services. The findings indicate that there exists a cognitive bias towards factors influencing the WLSQ. Crew members express the greatest concern over vessel lockage delays, whereas vessel lockage safety is the primary concern for management department administrators. Furthermore, enhancing the supporting facilities of waterway locks can significantly increase crew members' satisfaction during vessel lockage. Improving staff skills, and safety conditions can also greatly enhance customers' tolerance for lockage delays. The results of this study will provide valuable insights for the lock management department, operators, and the government in formulating relevant policies to improve WLSQ and implementing ongoing service quality evaluations.

**Highlights**

- ✓ This paper examines the factors that affect the waterway lock service quality (WLSQ) in the Yangtze River Delta.
- ✓ The WLSQ evaluation model is developed using factor analysis and structural equation model techniques.
- ✓ The study reveals cognitive biases in the perception of WLSQ between service suppliers and customers.
- ✓ The results indicate that vessel passing lock efficiency is the most significant factor influencing waterway lock service.

## 1. Introduction

In recent years, there has been a growing global focus on carbon emissions, leading to an increasing emphasis on reducing pollutant discharge across various modes of transportation (Strauss et al., 2021). Inland waterway transport is often lauded for its reliability, cost-effectiveness, and environmental friendliness, owing to its substantial transport capacity and low-carbon footprint (Jin et al., 2023; Li et al., 2023; Wang et al., 2023; Yang et al., 2023; Zhu et al., 2024). Additionally, it plays a crucial role in multimodal transportation networks, facilitating the international trade of goods to major hub ports. For instance, the inland waterway network in Jiangsu Province, particularly vital for the logistics of Shanghai Port, exemplifies this role. Against this backdrop, inland waterway locks are experiencing heightened activity. For instance, in 2020, the north section of the Beijing-Hangzhou Canal in Jiangsu province accommodated 1.32 million lockage vessels, transporting a total freight volume of 1.583 billion tons, with locks being opened a total of 279,600 times†. However, this increased activity has brought about various traffic-related challenges, including vessel delays and safety concerns. In certain waterway locks along the Yangtze River, vessel queuing delays have extended for several days (Deng et al., 2021; Tang, 2016; Wu and Lee, 2022). As Fig.1 shows, in Huai'an Lock, delays surged notably in 2021 compared to 2020, peaking in July with up to 5700 vessels queued in a single day. Moreover, safety issues during vessel lockage, such as collisions, groundings, and contact accidents, also pose ongoing risks.

---

* Corresponding author. *E-mail address*: pliao@seu.edu.cn; 1520686309@qq.com. *Tel.*: +86 25 5209 1265

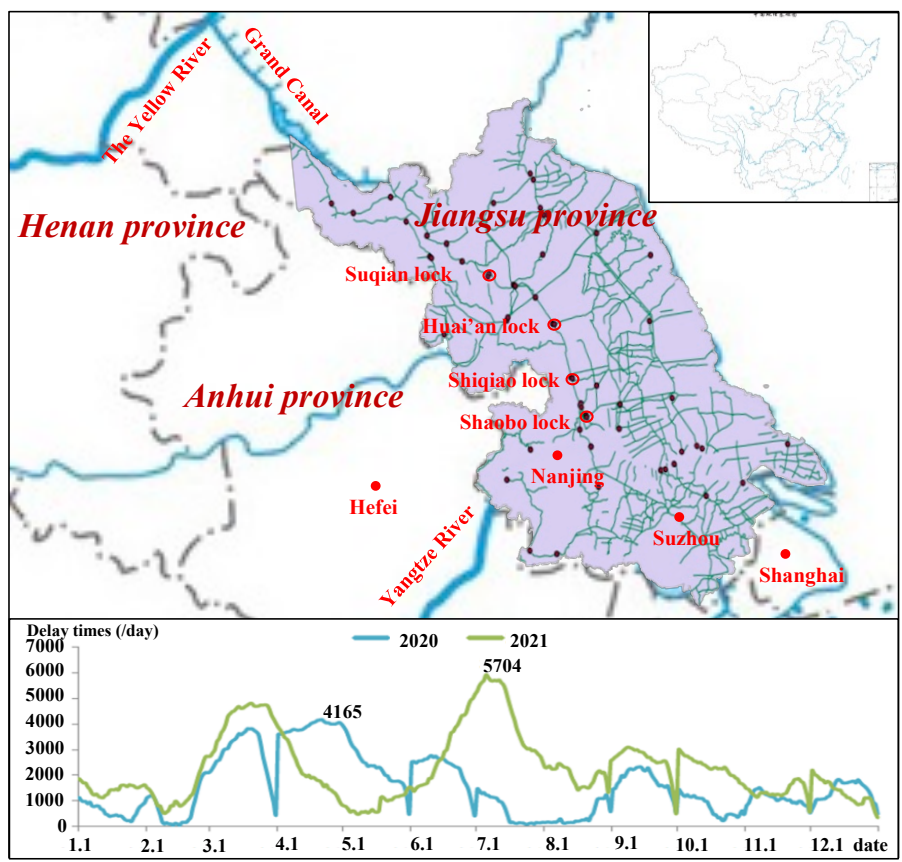

**Fig. 1.** Some typical waterway locks in Yangtze delta; Number of daily delayed vessels at Huai'an waterway lock (bottom).

The occurrence of these issues is intricately linked to the traffic dynamics of vessel lockage in the waterway locks. As depicted in Fig. 2, the inland waterway lock serves as a pivotal control point within the waterway network, facilitating the elevation and descent of vessels between low and high-level waterways (Liu et al., 2021; Wan et al., 2022; Wang and Li, 2022). As illustrated in Fig. 3, the primary sequence of events involved in vessel passage through a lock encompasses five distinct stages: declaration, payment, vessel queuing, dispatching, and lock passage. Initially, vessels arrive at the anchorage and formally declare their intention to traverse the lock. Subsequently, they receive notification of their scheduled passage time

---

† https://www.mot.gov.cn/jiaotonggaikuang/201804/t20180404_3006639.html (in Chinese, assessed on 2021 May. 20)

and are organized into a designated queue following completion of payment procedures. Upon completion, vessels proceed to the approach channel where they await release of lock passage instructions. Subsequently, scheduling queues are established, and vessels sequentially enter the lock. Ultimately, vessels exit the waterway lock upon completion of its operation. This entire process is characterized by its complexity and associated challenges, including significant vessel delays and inherent risks.

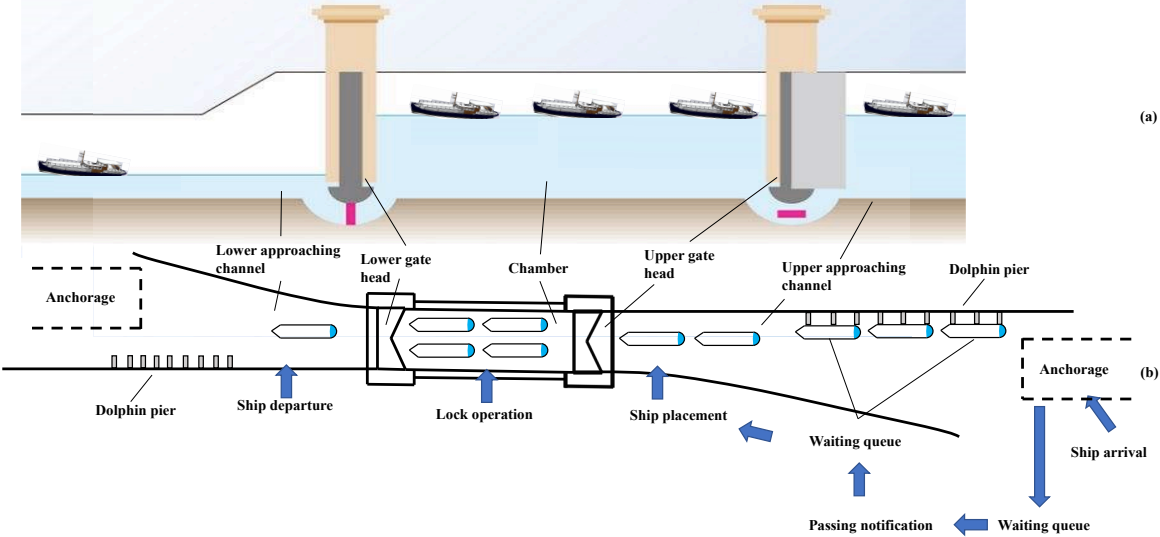

**Fig. 2.** Schematic diagram of inland waterway lock. (a) profile view, (b) vertical view.

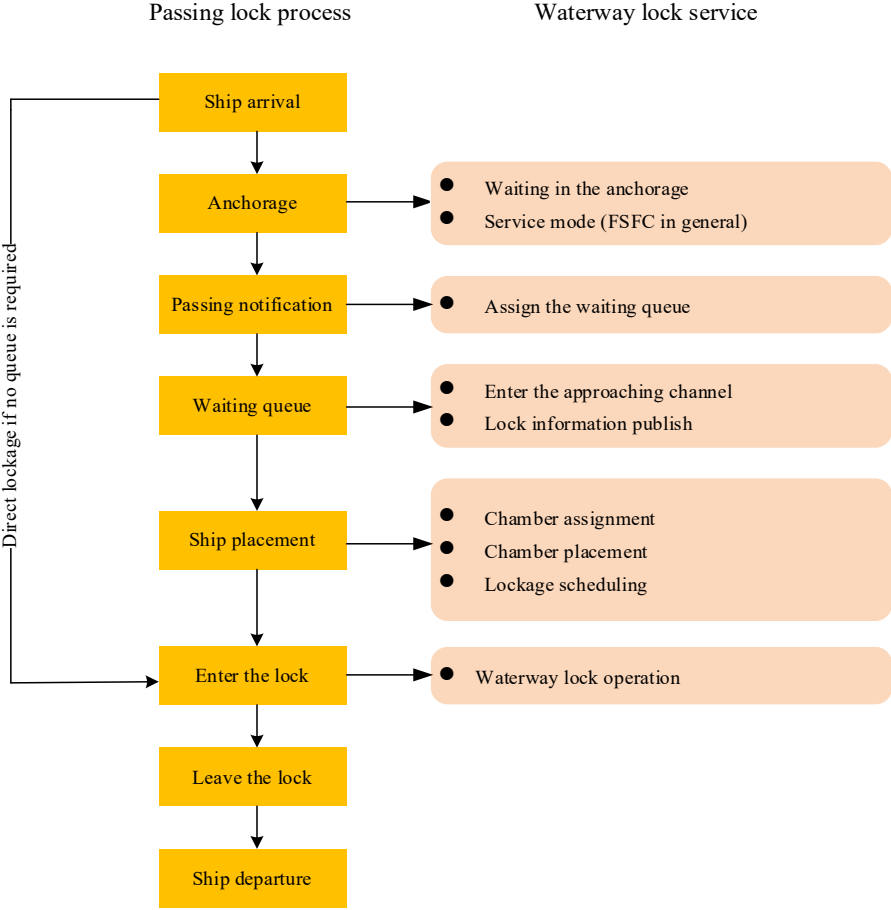

**Fig. 3.** Main process and corresponding services for vessel lockage.

The relevant departments responsible for waterway lock management have implemented several proactive measures aimed at enhancing waterway lock capacity and ensuring the safety of vessel lockage. Over the past decades, significant investments have been made in the construction and enhancement of waterway locks and associated infrastructure (Vooren et al., 2012; Liao, 2018). In China, for instance, additional chambers have been constructed alongside existing waterway locks to alleviate congestion. For example, a third chamber has been added to every eight locks in the Jiangsu segment of the Beijing-Hangzhou Grand Canal (Liao, 2018). Furthermore, vessel scheduling protocols have been optimized to streamline lockage procedures with advancements in waterway lock information technology (Liu et al., 2021; Deng et al., 2021; Cao et al., 2022). Collaborative scheduling strategies have been implemented for vessel passage through locks (Xia et al., 2021). These interventions have mitigated capacity constraints at waterway locks and yielded notable economic benefits.

The current interventions predominantly reflect the perspective of service suppliers. However, the quality of service experienced by customers has not received adequate consideration from suppliers. Hence, there is a necessity to assess customers' perceptions regarding existing delays and service standards during the locking process and identify their primary concerns. This is crucial for understanding potential conflicts of interest among various stakeholders, notably between crew members and management representatives. By integrating multiple factors to enhance the quality of lock services, we can significantly improve the satisfaction of crew members during lock passage.

In fact, there has been a growing interest in the service quality of transportation systems in recent years (Lee et al., 2016; Li and Yang, 2023; Chauhan et al., 2021; Khakdaman et al., 2022). Evaluating customer satisfaction has emerged as a pivotal metric for assessing the quality of transportation services (Li et al., 2013; Bivina et al., 2019; Chauhan et al., 2021). The bulk of this research has centered on urban public transport modalities such as bus transit (Dell' Olio et al., 2010), metro and rail transit (Vivek et al., 2022), and bicycles (Li et al., 2013; Bai et al., 2017). Researchers have predominantly explored the factors influencing service quality through customer satisfaction surveys and subsequently developed evaluation models using methodologies like Structural Equation Modeling (SEM), Analytical Hierarchy Process (AHP), ordered Probit models, among others. However, the majority of these studies have concentrated on public road transportation, with relatively little attention given to alternative perspectives.

Summarizing these research methods reveals several insights: (1) SEM proves invaluable for handling complex model structures, effectively modeling latent variables and measurement models, and supporting causal inference when analyzing questionnaire results. Therefore, SEM is well-suited for scrutinizing questionnaires with numerous items, with preliminary factor analysis of the original data being essential prior to modeling (Amadu et al., 2021; Chauhan et al., 2021; Wang et al., 2024). (2) The AHP, a method for multi-criteria decision analysis, offers systematic processing and analysis of questionnaire results. It proves particularly advantageous when comprehensive consideration of multiple factors and quantitative subjective judgments is required (Akdeniz et al., 2023; Chen et al., 2023; Zhen et al., 2023). (3) Ordered Probit models enable the evaluation of each factor's significance in outcomes, making them instrumental in simplifying factors (Bai et al. 2017; Vivek et al., 2022).

Therefore, this study aims to address the existing gap in waterway transportation and lock systems by integrating various service quality attributes. These attributes are crucial for evaluating service performance comprehensively. The overall Waterway Lock Service Quality (WLSQ) is assessed using factor analysis, SEM, and AHP methodologies. Factor analysis and SEM are applied to analyze customer questionnaires, while AHP is utilized for service supplier surveys. Additionally, we propose an assessment approach employing the ordered Probit method to ensure equitable evaluation of waterway lock service performance. The subsequent sections of this paper are structured as follows: Section 2 outlines the survey design and data collection process. Section 3 provides a brief overview of the statistical methods employed. In Section 4, we establish a service quality evaluation model for waterway locks and analyze the results from diverse viewpoints. Finally, Section 5 discusses the conclusions drawn from the study.

## 2. Research design and data collection

### 2.1. Research framework

Fig. 4 depicts the research framework employed in this study. The green rectangles represent the research methods utilized, while the yellow circles denote the research objectives or conclusions. To investigate the assessment of WLSQ from the perspectives of both customers and service providers, two sets of questionnaires were developed, one for customers and another for service providers, based on the observed variables outlined in Fig. 4. Factor analysis was conducted on the collected data from these questionnaires to identify the most significant factors influencing WLSQ. Subsequently, the latent variables in Fig. 4 were refined based on the results of the factor analysis. An AHP was then constructed using the refined variables to establish Model 2 for service providers. Additionally, SEM, denoted as Model 1, was employed to assess the impact of service quality attributes on the overall service quality of a waterway lock. By comparing Model 1 with Model 2, the cognitive bias of the waterway lock service was determined. Finally, a simplified model for the continuous evaluation of waterway lock service performance was developed using the ordered Probit method.

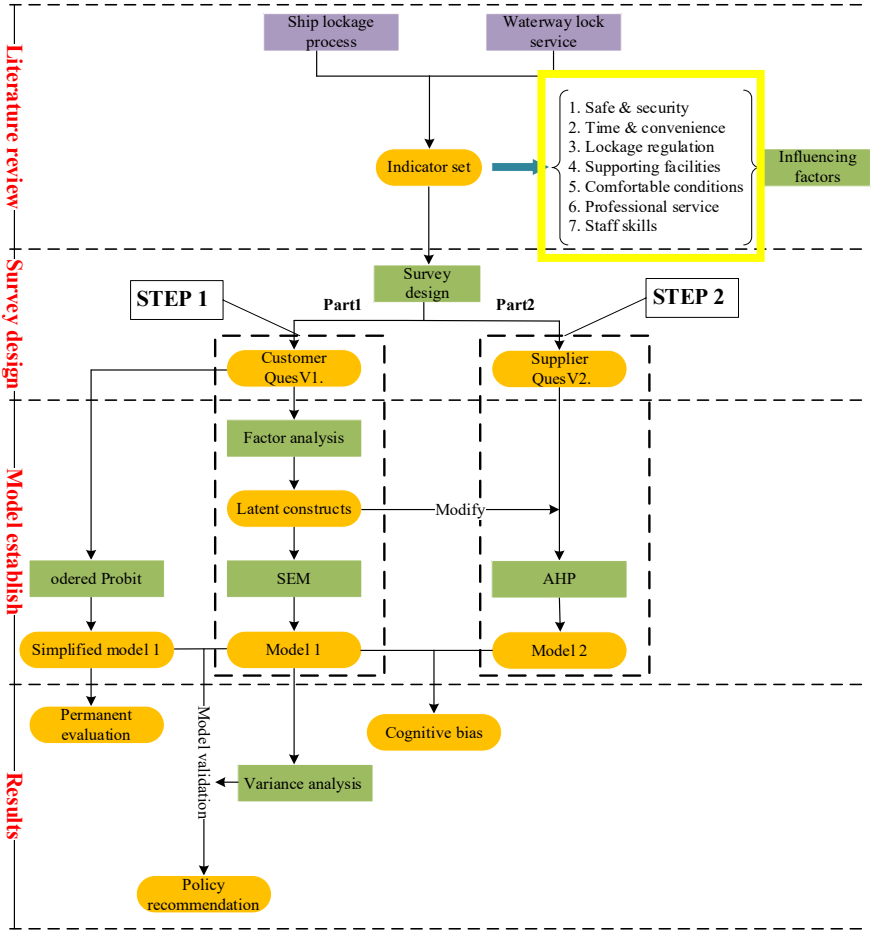

**Fig. 4.** Research framework.

## 2.2. Questionnaire design

### 2.2.1. Questionnaire for customer

The initial step in assessing the quality of transportation system services involves determining the observed variables. This selection process should be comprehensive, encompassing various aspects of transportation challenges (Eboli et al., 2007; Bai et al., 2017; Chauhan et al., 2021). For instance, when a vessel is awaiting lockage in the anchorage area or approaching the channel, the provision of safe, convenient, and comfortable waiting conditions plays a crucial role in alleviating crew anxiety and enhancing satisfaction with the lockage process. Additionally, the formulation of reasonable lockage scheduling rules and timely dissemination of information largely dictate the sequencing of lockages, directly impacting vessel delays and crew perceptions of waterway lock services (Wang et al. 2005; Verstichel et al, 2014; Deng et al., 2021; Liu et al., 2021).

As a result, operational efficiency, safety, regulatory performance, facility functionality, and service standards were identified as key dimensions for evaluating customer perceptions during lockage operations. These dimensions were further broken down into seven attributes specifically related to lockage: (1) Safe & security, (2) Time & convenience, (3) Lockage regulation, (4) Supporting facilities, (5) Comfortable conditions, (6) Service professional, (7) Staff skills. Subsequently, a total of thirty-two observed variables, outlined in Fig. 5, were proposed to gauge customer satisfaction with these dimensions.

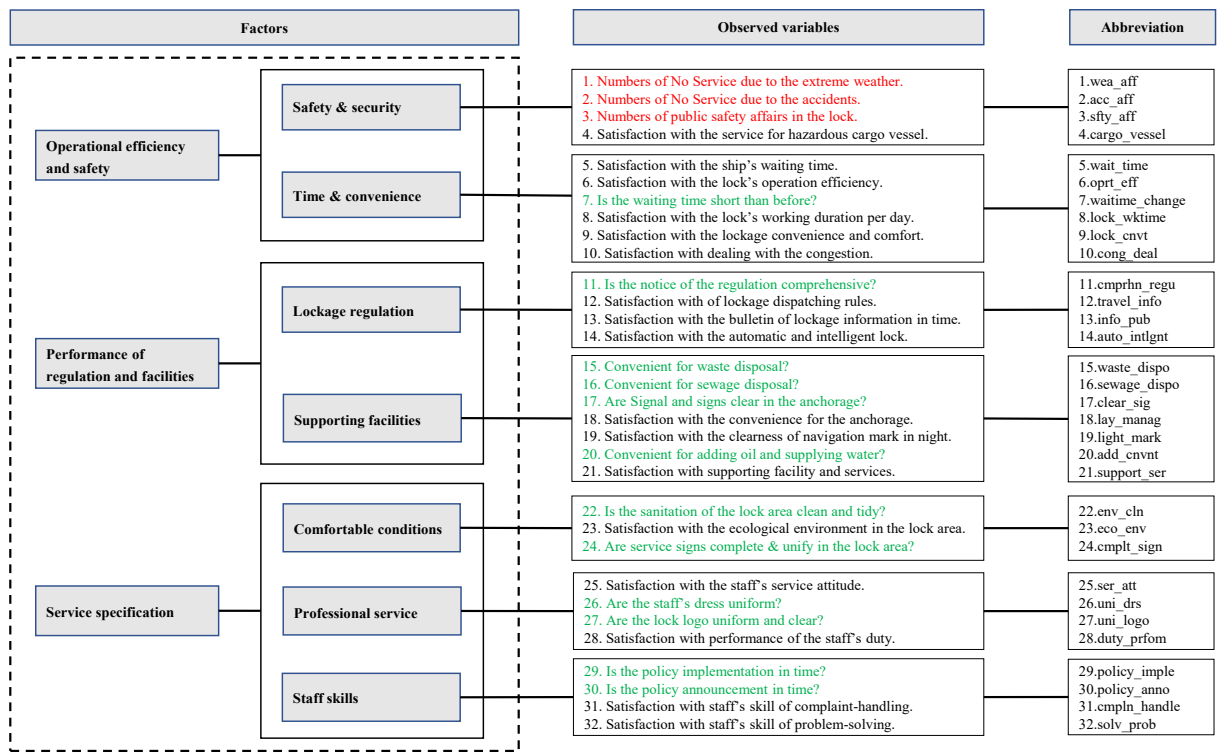

**Fig. 5.** Observed variables for waterway lock service from customer perception.

The questionnaire survey targeted crew members passing through the waterway locks in the Yangtze River Delta. The design rationale for the questions is detailed in Table 1, and a 5-point Likert scale was utilized to gauge customers' perceptions. Each participant was tasked with assessing their level of satisfaction across 32 observed variables, rating them on a Likert scale ranging from 1 (strongly dissatisfied) to 5 (strongly satisfied).

The final questionnaire comprised three sections: (1) Information concerning the vessel and crew, including demographics such as age, gender, years of experience, vessel type, and dead weight tonnage (DWT). (2) Ratings of satisfaction on a 5-point Likert scale for the 32 observed variables associated with WLSQ. (3) Crew members' suggestions regarding waterway lock services.

Moreover, drawing from prior research (Dell'Olil et al., 2010; Chauhan et al., 2021), it is acknowledged that respondents may alter their evaluation criteria upon becoming aware of various attributes affecting service quality. Therefore, assessments of overall service quality for waterway locks were conducted both at the beginning and conclusion of the survey, abbreviated as "sati_before" and "sati_after."

**Table 1.** Questions design basis.

| Setting aspects | Explanation | Example | Number of question |
|---|---|---|---|
| Frequency (Red font in Fig. 4) | Frequency of events | "Over the past year, how frequently have you encountered instances where locks were closed to traffic due to severe weather conditions?" | 1, 2, 3 |
| Subjective tendency (Green font in Fig. 4) | Customer's subjective perception of the service | "Do you perceive the current waiting time for vessels to be shorter compared to previous periods?" | 7, 11, 15~17, 20, 22, 24, 26~27, 29~30 |
| Satisfaction (Black font in Fig. 4) | Customer satisfaction rating of the service | "Are you satisfied with the problem-solving abilities of the waterway lock staff?" | remaining part |

*2.2.2. Questionnaire for service supplier*

In addition to the customer questionnaire, this study has also designed a separate questionnaire specifically for service suppliers. To ensure the uniformity of the evaluation framework, the questionnaire was structured based on the AHP methodology proposed by Saaty (1988). The latent variables identified through factor analysis (as detailed in section 4.1) served as the basis for comparing questionnaire items.

The survey targeted the management departments responsible for waterway locks along the Yangtze River in Jiangsu province, China. After providing clear explanations of the relevant factors, service suppliers were invited to assign scores indicating the importance of paired variables. These scores ranged from 1 to 5, allowing linguistic judgments (equally important, moderate important, important, very strong important, extremely important) to be translated into numerical values (Carrese et al., 2022). It is essential to note that the importance scores assigned to each latent variable must adhere to consistent criteria. The questionnaire used for assessing variable importance is provided in Appendix A.

*2.3. Descriptive analysis*

A pilot survey can serve as a valuable tool for assessing the complexity of a questionnaire and for refining any unclear or confusing questions based on feedback (Li et al., 2013). The results of the pilot survey indicated that customers were able to comprehend the questions clearly, and the majority spent between 3 to 30 minutes completing the survey, with over 95% falling within this range. This served as a benchmark for evaluating the questionnaire's validity. A total of 825 questionnaires were distributed to customers, of which 750 valid responses were received, representing a response rate of 90.9%. The overall Cronbach's Alpha coefficient was calculated to be 0.910, surpassing the threshold of 0.7, indicating strong reliability of the collected samples (Tavakol et al., 2011; Jomnonkwao et al., 2016).

The demographic analysis revealed that nearly all vessels surveyed operated within the Yangtze River

Delta. Approximately 70% of respondents fell within the middle-age bracket (between 30 and 50 years old), with more than 5 years of driving experience. Furthermore, the majority (86%) of vessels had a DWT of less than 1000 tons. Bulk cargo comprised the largest share of shipments, accounting for close to 50%, followed by energy materials such as coal, steel, and oil.

The satisfaction ratings for the 32 observed variables, along with their overall service quality ratings, are depicted in Fig. 6. Utilizing a Likert scale ranging from 1 to 5, satisfaction ratings indicated a high level of satisfaction among crew members with the overall WLSQ, averaging nearly 4 out of 5. However, variables related to Time & convince and Supporting facilities scored lower, with higher standard deviations. In contrast, customers expressed greater satisfaction with variables associated with Safe & security and Staff skills. To assess the normal distribution of the data, skewness and kurtosis of the satisfaction ratings were analyzed. Following guidelines proposed by Muthén et al. (1985), skewness and kurtosis values close to zero, with an acceptable range between -1.50 and 1.50, indicate good normality of the data, as illustrated in Fig. 6. Subsequently, a structural equation model was employed to further analyze the data.

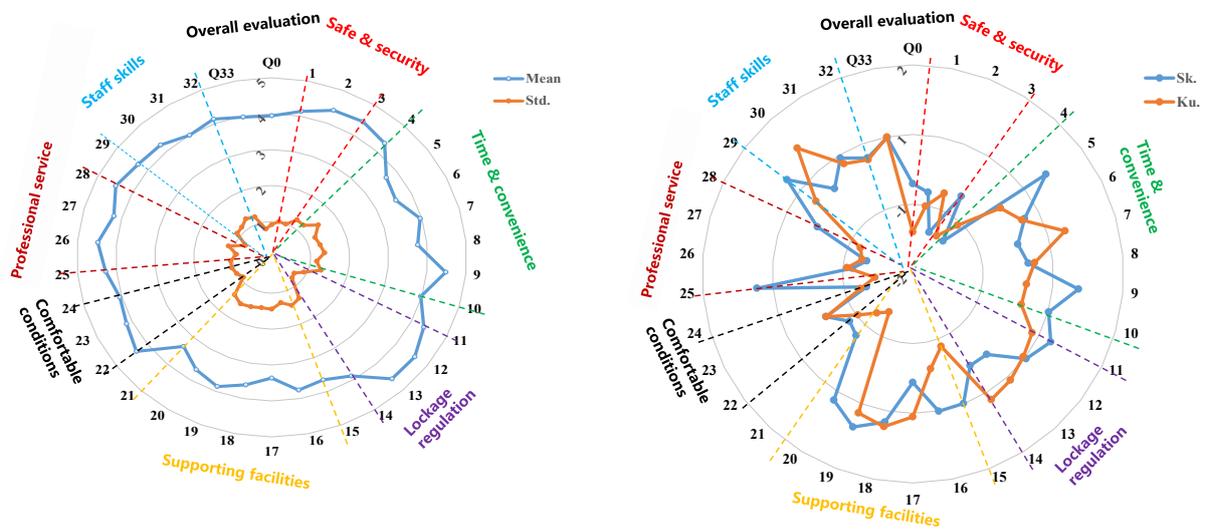

**Fig. 6.** Descriptive statistics of observed variables (satisfaction ratings, N=750).
**Note:** Std. = Standard Deviation; Sk. = Skewness; Ku. = Kurtosis; No. of variables are shown in Fig. 5; Q0 and Q33 are sati_before and sati_after, respectively.

## 3. Methodology

### 3.1. Factor analysis

Factor analysis is a statistical technique used to explore the hypothetical constructs underlying observed variables, aiming to identify correlation patterns among these variables (Raykov et al, 2006; Li et al., 2013; Jomnonkwao et al., 2016). It encompasses both exploratory factor analysis (EFA) and confirmatory factor analysis (CFA). In this study, the purpose of employing factor analysis is to uncover latent variables that describe WLSQ, based on the observed variables depicted in Fig. 5, while preserving the explanatory capacity of the original observed variables.

### 3.2. SEM

SEM is a versatile statistical technique widely employed in research for examining the relationships among multiple variables. It allows for the exploration of connections between both endogenous and exogenous variables, along with latent variables that are not directly observable but are represented as linear combinations of observable variables. This approach offers a powerful framework for understanding complex systems (Washington et al., 2010; Li et al., 2013; Chauhan et al., 2021).

In contexts where certain variables, such as service quality, are not directly observable but can be inferred from other measurable factors, SEM provides a valuable tool. For instance, in the study of waterway lock services, where service quality is a crucial but intangible aspect, SEM allows researchers to construct detailed models using observed variables to capture and analyze these underlying dynamics.

SEM employs various methods for covariance analysis, with maximum likelihood estimation being the most commonly used due to its robustness and efficiency (Bivina et al., 2019; Chauhan et al., 2021). This methodological choice ensures the accuracy and reliability of the model in capturing the intricate relationships among the observed variables and their latent constructs.

*3.3. AHP*

The AHP is a method commonly employed to address selection dilemmas and is extensively utilized in multiple criteria decision-making scenarios. This technique enables decision-makers to systematically break down complex decisions into constituent parts, assigning weights and rankings to each aspect of the problem (Saaty, 1988). In this study, AHP was employed to both design and analyze a questionnaire aimed at assessing service suppliers.

*3.4. Ordered Probit*

The ordered probability model emerges as a suitable framework for capturing ordered discrete data, as highlighted by prior research (Washington et al., 2003; Bai et al., 2017). In this study, we propose the development of an ordered Probit model to streamline customer questionnaires and facilitate continuous assessment of the WLSQ. Given that the variables under consideration are characterized by an ordered discrete nature, the utilization of the ordered Probit model is warranted to streamline their representation.

**4. Results and discussions**

*4.1. Factor analysis results*

*4.1.1. EFA results*

The EFA was initially used to identify the appropriate set of latent factors underlying the 32 original observed variables depicted in Fig. 5. A total of 750 samples were selected and randomly divided into two sets. Factor analysis and SEM were performed on 450 samples, while the remaining 300 samples were reserved for model validation.

Prior to analysis, Kaiser-Meyer-Olkin (KMO) and Bartlett's test of sphericity were conducted. The KMO value was found to be 0.874, exceeding the recommended threshold of 0.6, indicating adequacy for factor analysis (Kaiser, 1960). Additionally, the significance level of Bartlett's test was below 0.01 (Kaiser, 1960), further supporting the suitability of the data for factor analysis.

In this research, the EFA was conducted using SPSS 26.0, focusing on satisfaction ratings for the 32 observed variables. Principal component analysis (PCA) was employed for factor extraction, retaining only factors with eigenvalues greater than 1 (Guttman, 1954). Observed variables with factor loadings below 0.5 were eliminated (Hair et al., 2014).

Following convergence, the rotated component matrix revealed that three observed variables (No. 4, 27,

28) were removed due to low factor loadings or cross-loading on multiple factors. Ultimately, 29 variables were retained, revealing six latent factors: (a) Supporting facilities (No.14-20), (b) Staff skills (No.24-29), (c) Time & convince (No.4-9), (d) Lockage regulation (No.10-13), (e) Comfortable conditions (No.21-23), (f) Safe & security (No.1-3). These factors collectively accounted for 63.05% of the total variance. It's worth noting that the original category of lock specification (Fig. 5) was clustered with work execution, as the distinction between problem design and execution was not clearly delineated. Table 2 presents the results of the EFA, including the latent factors, the corresponding observed variables, factor loadings, variance explained, and Cronbach's alpha coefficients.

Table 2. Results of EFA. (The meaning of observed variables is described in Fig. 5.)

| Latent factor | Observed variables | Factor loading | Variance Explained (%) | Cronbach Alpha |
|---|---|---|---|---|
| Time & convince | (9) lock_cnvt | 0.839 | 25.566 | 0.871 |
| | (5) wait_time | 0.834 | | |
| | (8) lock_wktime | 0.805 | | |
| | (10) cong_deal | 0.782 | | |
| | (6) oprt_eff | 0.760 | | |
| | (7) waitime_change | 0.728 | | |
| Supporting facilities | (18) lay_manag | 0.767 | 11.734 | 0.859 |
| | (20) add_cnvnt | 0.764 | | |
| | (15) waste_dispo | 0.732 | | |
| | (19) light_mark | 0.722 | | |
| | (16) sewage_dispo | 0.677 | | |
| | (17) clear_sig | 0.667 | | |
| | (21) support_ser | 0.660 | | |
| Staff skills | (32) solv_prob | 0.825 | 8.674 | 0.838 |
| | (31) cmpln_handle | 0.770 | | |
| | (25) ser_att | 0.694 | | |
| | (29) policy_imple | 0.650 | | |
| | (30) policy_anno | 0.645 | | |
| | (26) uni_drs | 0.592 | | |
| Lockage regulation | (13) info_pub | 0.827 | 6.582 | 0.779 |
| | (12) travel_info | 0.768 | | |
| | (11) cmprhn_regu | 0.760 | | |
| | (14) auto_intlgnt | 0.632 | | |
| Comfortable conditions | (23) eco_env | 0.852 | 5.680 | 0.826 |
| | (24) cmplt_sign | 0.804 | | |
| | (22) env_cln | 0.731 | | |
| Safe & security | (2) acc_deal | 0.863 | 4.814 | 0.784 |
| | (1) wea_deal | 0.802 | | |
| | (3) inc_deal | 0.795 | | |
| Cumulative Explained Variance | | | 63.050 | |

*4.1.2. CFA results*

CFA was conducted using AMOS 24.0, building upon the outcomes of EFA. Table 3 presents the results of the first-order measurement model assessment. The composite reliability (C.R.) values range from 0.79 to 0.90, surpassing the threshold of 0.7 as suggested by Fornell et al. (1981). This indicates a notable level of internal consistency among the observed variables. Furthermore, the average variance extracted (AVE) of the most latent variables exceeds 0.5. The standardized regression weights (Std) of each observed variable are above 0.6, confirming the convergent validity of the constructs. Additionally, regression weights (Unstd), standard errors (SE), t-values, and p-values (less than 0.001) were calculated to ascertain the significance of these latent variables.

**Table 3.** Result of first-order measurement model assessment.

| Construct | Variables | Unstd. | S.E. | t-value | p | Std. | SMC | CR | AVE |
|---|---|---|---|---|---|---|---|---|---|
| Safe & security | 1 | 1.000 | | | | 0.759 | 0.576 | 0.796 | 0.567 |
| | 2 | 1.065 | 0.083 | 12.863 | *** | 0.827 | 0.684 | | |
| | 3 | 1.063 | 0.086 | 12.393 | *** | 0.665 | 0.442 | | |
| Time &convince | 4 | 1.000 | | | | 0.876 | 0.767 | 0.895 | 0.588 |
| | 5 | 0.866 | 0.059 | 14.750 | *** | 0.731 | 0.534 | | |
| | 6 | 0.919 | 0.059 | 15.703 | *** | 0.667 | 0.445 | | |
| | 7 | 0.956 | 0.049 | 19.364 | *** | 0.763 | 0.582 | | |
| | 8 | 0.862 | 0.043 | 19.972 | *** | 0.808 | 0.653 | | |
| | 9 | 0.893 | 0.050 | 17.792 | *** | 0.740 | 0.548 | | |
| Lockage regulation | 10 | 1.000 | | | | 0.728 | 0.530 | 0.821 | 0.538 |
| | 11 | 0.821 | 0.058 | 14.124 | *** | 0.760 | 0.578 | | |
| | 12 | 0.970 | 0.065 | 14.991 | *** | 0.846 | 0.716 | | |
| | 13 | 1.136 | 0.101 | 11.285 | *** | 0.573 | 0.328 | | |
| Supporting facilities | 14 | 1.000 | | | | 0.700 | 0.490 | 0.851 | 0.451 |
| | 15 | 0.924 | 0.071 | 13.022 | *** | 0.722 | 0.521 | | |
| | 16 | 0.864 | 0.078 | 11.118 | *** | 0.608 | 0.370 | | |
| | 17 | 0.961 | 0.078 | 12.305 | *** | 0.654 | 0.428 | | |
| | 18 | 0.906 | 0.081 | 11.219 | *** | 0.597 | 0.356 | | |
| | 19 | 1.100 | 0.083 | 13.256 | *** | 0.765 | 0.585 | | |
| | 20 | 0.896 | 0.080 | 11.259 | *** | 0.636 | 0.404 | | |
| Comfortable conditions | 21 | 1.000 | | | | 0.743 | 0.552 | 0.829 | 0.619 |
| | 22 | 1.205 | 0.083 | 14.574 | *** | 0.767 | 0.588 | | |
| | 23 | 1.250 | 0.085 | 14.776 | *** | 0.846 | 0.716 | | |
| Staff skills | 24 | 1.000 | | | | 0.757 | 0.573 | 0.828 | 0.448 |
| | 25 | 0.696 | 0.065 | 10.765 | *** | 0.573 | 0.328 | | |
| | 26 | 0.796 | 0.069 | 11.521 | *** | 0.624 | 0.389 | | |
| | 27 | 0.804 | 0.069 | 11.661 | *** | 0.617 | 0.381 | | |
| | 28 | 0.966 | 0.077 | 12.606 | *** | 0.667 | 0.445 | | |
| | 29 | 1.048 | 0.074 | 14.193 | *** | 0.756 | 0.572 | | |

**Note:** Unstd. = Regression Weight; S.E. = Standard Error; t-val = t-value; Std. = Standardized Regression Weight; SMC = Square Multiple Correlations (Item Reliability); C.R. = Composite Reliability; AVE = Average Variance Extracted; "***" denotes p-value ≤ 0.001.

The discriminant validity of the constructs was assessed using the Fornell-Larcker criterion (Fornell et al., 1981). This criterion involves comparing the square root of the AVE for each construct with the Pearson correlation coefficients between pairs of constructs, as outlined in Table 4. The diagonal values in the table consistently surpass the Pearson correlation coefficients in their respective rows and columns. This pattern confirms the presence of discriminant validity among the constructs.

**Table 4.** Discriminant validity analysis of the first-order model.

| | AVE | Comfortable conditions | Supporting facilities | Lockage regulation | Time & convince | Staff skills | Safe & security |
|---|---|---|---|---|---|---|---|
| Comfortable conditions | 0.619 | **0.787** | | | | | |
| Supporting facilities | 0.451 | 0.321 | **0.672** | | | | |
| Lockage regulation | 0.538 | 0.465 | 0.424 | **0.733** | | | |
| Time & convince | 0.588 | 0.213 | 0.199 | 0.308 | **0.767** | | |
| Staff skills | 0.448 | 0.588 | 0.463 | 0.562 | 0.262 | **0.669** | |
| Safe & security | 0.567 | 0.146 | 0.231 | 0.298 | 0.276 | 0.272 | **0.753** |

*4.2. SEM establishment and validation*

*4.2.1. Model establishment*

The SEM used to measure WLSQ with standard regression weights is illustrated in Fig. 7. This model aims to estimate the relationship between WLSQ and various observed variables associated with customer satisfaction. Model fit indices, presented in Table 5, all meet the specified criteria. The results of the SEM estimation can be found in Table 6. The coefficients between the latent factors and observed variables, as estimated in the SEM, align with the CFA results. Notably, all variance estimates are greater than 0, with standard deviations ranging from 0.179 to 0.417. The critical ratio (C.R.) of parameter estimates and standard errors exceeds 1.96, indicating statistical significance at the 0.001 level (Chauhan et al., 2021). These indices collectively suggest that the final SEM model developed in this study exhibits a strong fit.

**Table 5.** Model fit indices.

| Model fit index | Value | Acceptable value | Source |
|---|---|---|---|
| CMIN/DF | 1.829 | Less than 3 | Carmines et al., 1981 |
| RMSEA | 0.043 | Less than 0.08 | Lomax et al., 2004 |
| CFI | 0.948 | | Bentler et al., 1990 |
| GFI | 0.901 | | Tanaka et al., 1985 |
| AGFI | 0.880 | Higher than 0.8 | Jackson et al., 2009 |
| NFI | 0.894 | | Bentler et al., 1980 |
| TLI | 0.942 | | Bollen, 1989 |
| IFI | 0.949 | | |

**Note:** CMIN/DF = Chi-square/degrees of freedom; RMSEA = Root-mean-square error of approximation; CFI = Comparative fit index; GFI = Goodness of fit index; AGFI = Adjusted goodness of fit index; NFI = Normed fit index; TFI = Tucker–Lewis index; IFI = Incremental fit index.

**Table 6.** Estimate results of SEM.

| Path direction | Latent factor | Unstd. | S.E. | C.R. | p | Std. |
|---|---|---|---|---|---|---|
| | Safe & security | 0.200 | 0.033 | 6.070 | *** | **0.235** |
| | Time & convince | 0.277 | 0.026 | 10.778 | *** | **0.417** |
| Service quality | Lockage regulation | 0.147 | 0.042 | 3.471 | *** | **0.151** |
| | Supporting facilities | 0.201 | 0.030 | 6.623 | *** | **0.300** |
| | Comfortable conditions | 0.171 | 0.043 | 4.017 | *** | **0.179** |
| | Staff skills | 0.188 | 0.040 | 4.685 | *** | **0.247** |

**Note:** Unstd. = Regression Weight; S.E. = Standard Error; Std. = Standardized Regression Weight; C.R. = Critical Ratio; "***" denotes p-value ≤ 0.001.

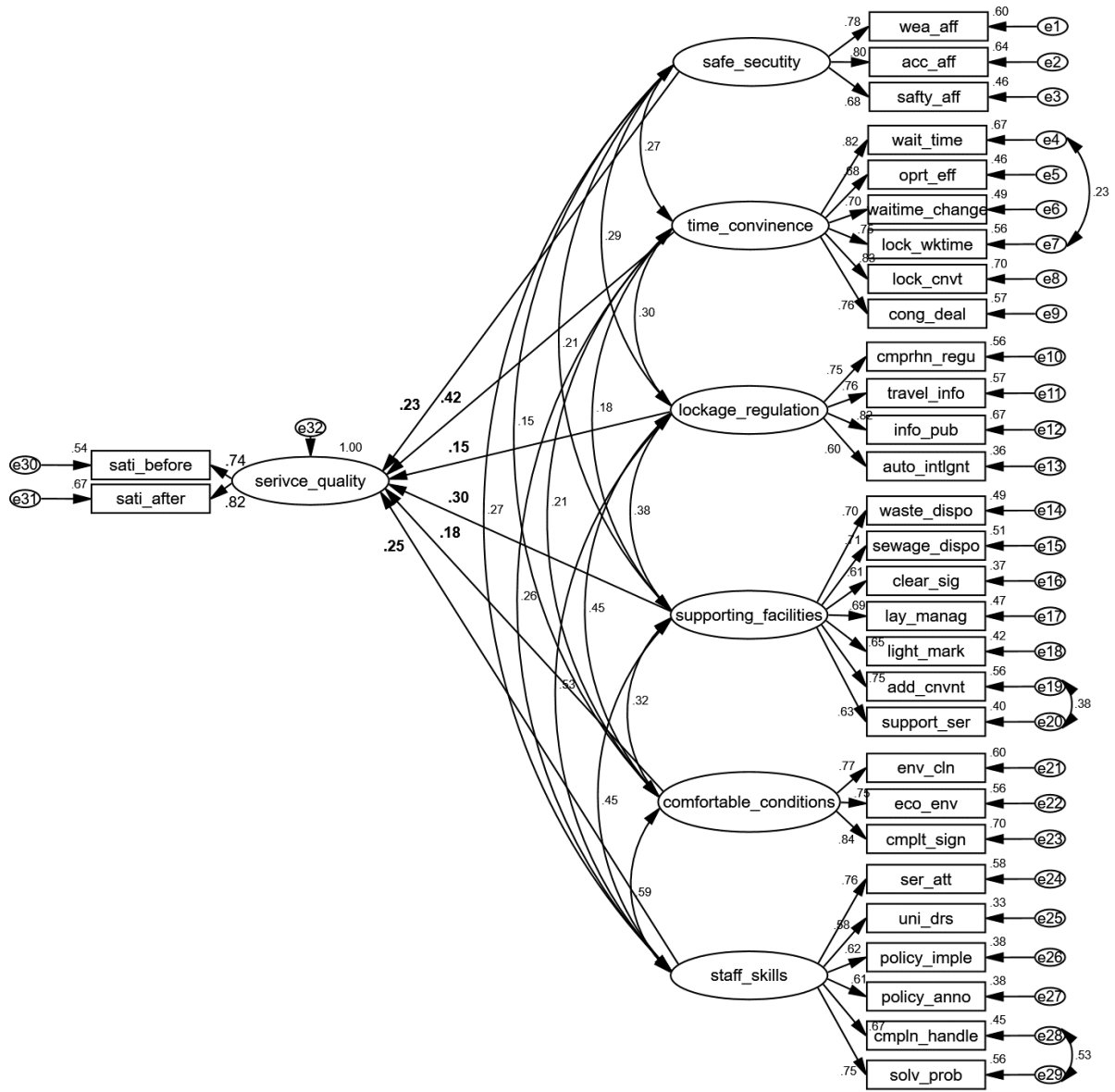

**Fig. 7.** Structural equation model for measuring WLSQ.

*4.2.2. Model validation*

Model validation assesses the model's capability to accurately estimate the expected level of fit. A validated model yields more precise prediction results (James et al., 2013; Chauhan et al., 2021). In this study, the WLSQ evaluation model was developed using 450 samples, with the remaining 300 data points allocated for model validation. The satisfaction rating of latent variables (*LVR*) was computed following Eq. 1, while the overall evaluation rating of WLSQ ($\widehat{SQR}$) was determined using Eq. 2. The disparity between the satisfaction rating at the conclusion of the survey (sati_after) and $\widehat{SQR}$ was quantified via Eq. 3, and the outcomes of model validation are depicted in Fig. 8. The error ranges are represented using gradient colors, with the median error of each satisfaction rating depicted as a solid circle. Notably, as Fig. 8 shows, the majority of errors (approximately 82%) fall within the range of -0.1 to 0.1. Additionally, the mean error falls within ±7.91%, indicating a prediction accuracy of less than 10% for this model.

$$LVR_i = \frac{\sum_{j=1}^{m}(OVR_{i,j} \times SRW_{i,j})}{\sum_{j=1}^{m}SWR_{i,j}} \tag{1}$$

where $LVR_i$ represents the rating of the $i$-th latent variable. $OVR_{i,j}$ denotes the satisfaction rating of the $j$-th observed variable with respect to the $i$-th latent variable. $SRW_{i,j}$ represents the standardized regression weight of the $i$-th latent variable on the $j$-th observed variable. $m$ signifies the number of observed variables associated with the $i$-th latent variable.

$$\overline{SQR}_x = \frac{\sum_{i=1}^{n}(LVR_i \times SRW_i)}{\sum_{i=1}^{n}SRW_i} \tag{2}$$

where $\overline{SQR}_x$ refers to the WLSQ rating of the $x$-th respondent estimated by model. $SRW_i$ represents the standardized regression weight associated with the $i$-th latent variable, as outlined in Table 6. $n$ denotes the total number of latent variables considered.

$$error = \frac{|sati\_after - \overline{SQR}|}{sati\_after} \tag{3}$$

Moreover, as depicted in Fig. 8, a significant portion of the respondents provided a low overall satisfaction rating for WLSQ, despite assigning relatively high ratings to specific observed variables. This phenomenon arises from customers' tendency to favor particular observed variables, such as delay, supplement, and service attitude, over others when assessing overall satisfaction. Consequently, their evaluations may overlook the impact of other factors. This underscores the limitations of relying solely on a single perspective for evaluation.

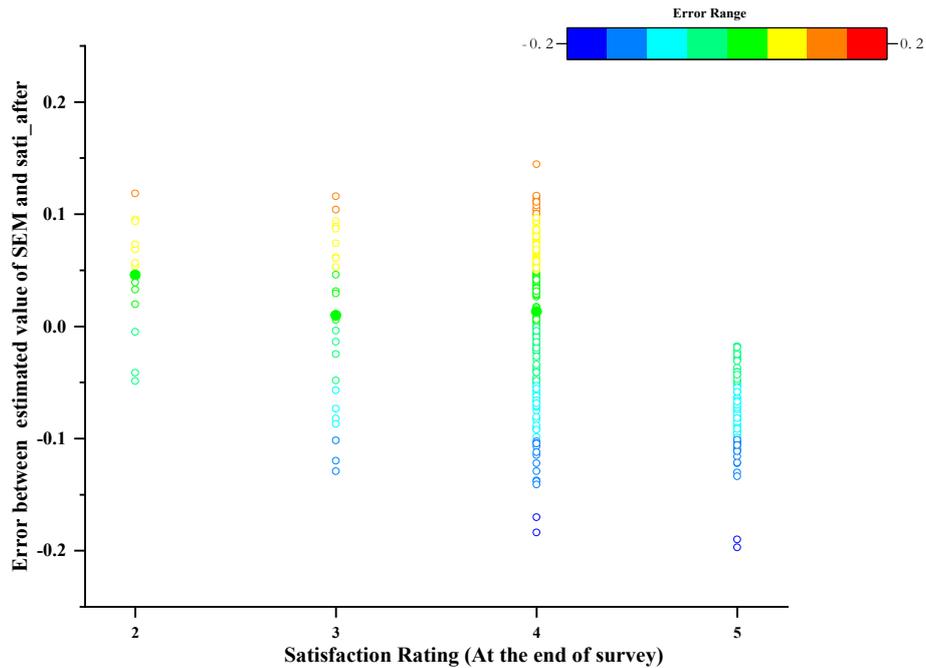

**Fig. 8.** Model validation result.

*4.3. Model analysis from customer's perspective*

*4.3.1. Overall description*

Based on the findings from Figure 7 and Table 6, it is evident that "Time & convenience" (γ = 0.417, p-value<0.001) and "Supporting facilities" (γ = 0.300, p-value<0.001) have the most significant impact on WLSQ from the perspective of crew members. Following closely are "Staff skills" (γ = 0.247, p-value<0.001) and "Safe & security" (γ = 0.235, p-value<0.001). Additionally, "Comfortable conditions" (γ = 0.179, p-value<0.001) and "Lockage regulation" (γ = 0.151, p-value<0.001) moderately influence WLSQ. This can be attributed to the high delays often experienced during the vessel lockage process (Verstichel et al, 2014; Liu et al., 2021; Deng et al., 2021), which is a major concern for customers. Therefore, the availability of supporting facilities such as supplies, lighting, and waste disposal during waiting times plays a critical role and is reflected in the latent variable of "Supporting facilities".

Furthermore, the results indicate that customer satisfaction is influenced by the performance of lock management staff in service execution, including their attitude and efficiency in problem-solving. However, customers do not assign significant importance to the environment or intelligence level of waterway locks. This is because their main concern lies in the accuracy and timely release of information for crossing locks, and the current information release method and operation mode of the locks are deemed satisfactory.

The study also reveals that the standardized regression weight of sati_before (γ = 0.735) is lower than that of sati_after (γ = 0.822), indicating that respondents evaluated service quality more positively after being informed about the attributes of WLSQ. This finding aligns with previous research conducted by Dell'Olil et al. (2010) and Chauhan et al. (2021).

*4.3.2. Customer satisfaction across various delay levels*

Delays during vessel lockage, particularly during peak traffic periods, are often inevitable. By providing better services to customers during these delay periods, it can help reduce anxiety and improve their perception of service satisfaction. However, it is important to determine which enhancements in waterway lock services can significantly improve overall service performance. To achieve this, it is necessary to classify customers based on the duration of delays and analyze the variability of customer satisfaction at different levels of delay. Additionally, it is crucial to discuss sensitive indicators that reflect higher customer expectations during delays (Roucolle et al., 2020).

Based on the results of the vessel waiting time, the 750 valid data collected were divided into various delay intervals. The mean score of question 33 (sati_after) within each interval was used to represent the customers' overall rating value. To calculate the coefficient of variability of crew members' satisfaction with the observed variable *i* at different levels of delay, the information entropy method is applied (Liu et al., 2010):

$$E_i = -\frac{1}{\ln N} \sum_{k=1}^{N} P_{ik} \ln(P_{ik}) \tag{4}$$

$$P_{ik} (= y_{ik} / \sum_{k=1}^{N} y_{ik}) \tag{5}$$

where $E_i$ ($0 \leq E_i \leq 1$) represents the information entropy of the *i*-th latent variable. $P_{ik}$ denotes the proportion of the *i*-th observed variable of the *k*-th latent variable. $N$ is the total number of valid questionnaires in the study.

When analyzing customer satisfaction values at different delay levels, variables related to the latent variable of "Time & Convenience" (No. 4-9) were excluded. The combined evaluation value ($S$) for each delay level was calculated and is shown in Table 7. It is evident that customer satisfaction increases as the delay time decreases. For delays within 2 hours, the satisfaction level is 4.45 (89%), while it drops to 2.13 (42.6%) for vessel delays exceeding 16 hours. Additionally, the overall satisfaction of crew members (3.97) falls between the ranges of 2-8 hours. This finding aligns with the current lock operation situation in Jiangsu province, China (Kong et al., 2017).

**Table 7.** The combined evaluation value of each delay level.

| Delay(h) | [0,2] | [2,4] | [4,8] | [8,16] | >16 |
|---|---|---|---|---|---|
| Percentage (%) | 37.1 | 27.0 | 11.2 | 13.5 | 11.2 |
| S | 4.45 | 4.12 | 3.35 | 2.84 | 2.13 |

The variability coefficients of each latent variable are presented in Table 8. Notably, "Supporting facilities", "Staff skills", and "Safe & security" exhibit substantial variability. This is closely followed by "Lockage regulation" and "Comfortable conditions". The findings suggest that customers with varying levels of delay have notably distinct expectations regarding the aforementioned three latent variables. Hence, enhancing "Supporting facilities", "Staff skills", and "Safe & security" can substantially enhance customers' ability to tolerate lockage delays.

**Table 8.** The coefficients of variability for each indicator.

| Latent variable | Safe & security | Lockage regulation | Supporting facilities | Comfortable conditions | Staff skills |
|---|---|---|---|---|---|
| E | 0.995 | 0.998 | 0.990 | 1.000 | 0.992 |
| $(1-E)/10^{-3}$ | 7.148 | 2.303 | 9.565 | 0.049 | 7.614 |

*4.4. Cognitive bias analysis and policy recommendation*

In this study, a survey was conducted on the management departments of forty-nine waterway locks in Jiangsu province, China. A total of 43 valid questionnaires were collected, representing 87.7% of the surveyed departments. The subjective weight, calculated as the geometric mean of the scores for each potential variable, reflected the perspective of the service suppliers. To represent the customer perspective, the standardized regression weights of the latent variables in Table 6 were normalized to obtain the objective weight.

**Table 9.** Weight of each factor from different perspectives.

| | Safe & security | Time & convince | Lockage regulation | Supporting facilities | Comfortable conditions | Staff skills |
|---|---|---|---|---|---|---|
| Criteria layer | Waterway lock operation efficiency | | Waterway lock function performance | | Waterway lock management specification | |
| OW (Customer) | 0.154(4) | **0.273**(1) | 0.099(6) | **0.196**(2) | 0.117(5) | **0.162**(3) |
| SW (Service supplier) | **0.244**(1) | 0.144(4) | **0.145**(3) | 0.109(6) | 0.155(5) | **0.203**(2) |

**Note:** OW = objective weight; SW = subjective weight; Number in () represents the ranking; The numbers in bold means the more important latent variable in the same criteria layer.

The weight of each factor from diverse perspectives is summarized in Table 9. It is evident that crew members, who are the customers, give priority to timely passage and reduced time costs. Conversely, waterway lock management departments, which are the service suppliers, prioritize safe and secure passage through the locks. Furthermore, service providers attach greater importance to Lockage regulations than supporting facilities due to the crucial nature of intelligent transportation for inland waterway transportation and lockage safety. However, customers prioritize the condition of the waterway lock, such as convenience and time cost, over intelligence.

Based on the findings of this study, the following policy recommendations are proposed:

Firstly, customers are primarily concerned with time and convenience, and therefore optimizing vessel lockage efficiency is essential. This can be achieved by implementing vessel scheduling rules that address congestion and simplifying the procedure for crossing waterway locks. Additionally, value-added services should be enhanced at locks, and delay information should be readily available in real-time to customers. Suppliers, on the other hand, prioritize safe and secure vessel passage, which can be achieved through strengthening emergency response to accidents and addressing situations where waterway locks are blocked in a timely manner.

Secondly, both customers and suppliers value staff skills, with active handling of issues and prompt feedback significantly enhancing waterway lock service performance. Therefore, it is recommended that staff training and development programs be implemented to improve these skills.

Thirdly, supporting facilities such as basic amenities, shore power, leisure facilities, communication options, and environmental protection for vessels should be improved to enhance the overall customer and supplier experience. Furthermore, alternative services based on the flow of vessels passing through different locks should be offered.

Lastly, while lockage regulation and comfortable conditions may have less influence on customer and supplier satisfaction, they should still be addressed through accurate and timely dissemination of information, optimized vessel scheduling methods, and maintenance of comfortable conditions.

*4.5. Simplification of observed variables*

The current service elevation system is not suitable for permanent assessment due to the large number of observed variables. However, considering that service satisfaction is an ordered hierarchical variable and observed variables are categorical, the ordered Probit model was chosen to simplify the evaluation system. Explanatory variables 1 to 29 were used in the model, with the overall service quality rating of the waterway lock as the explained variable. The ordered Probit analysis was conducted using STATA 15.1, and the results are presented in Table 10.

Model 1 is composed of twenty-nine explanatory observed variables, and its chi-square test indicates a good overall fit. Among the observed variables, those bolded in the table have a significant effect on customer satisfaction ($p<0.01$), while variables 21 and 23 have a negligible effect on crew satisfaction ($p>0.5$). The remaining variables display a moderate effect on customer satisfaction. Model 2 was created by excluding insignificant variables, resulting in higher variable significance compared to Model 1. The goodness of fit improved in terms of the log-likelihood ratio and pseudo coefficient of determination.

Based on the analysis of customers' satisfaction and the results of the ordered Probit model, the following conclusions can be drawn:

(1) Out of the 29 variables involved in ordered Probit Model 1, only 11 variables had significant effects on customer satisfaction. After re-running ordered Probit Model 2 with only these 11 variables, the goodness of fit of the model remained stable. Thus, the customers' satisfaction model can be simplified to these 11 variables.

(2) Two latent variables, "Lockage regulation" and "Comfortable conditions", have a minor effect on customer satisfaction during vessel lockage. Therefore, it is not recommended to analyze these two factors using a questionnaire service.

Finally, a simplified questionnaire consisting of the 11 observed variables highlighted in Table 10 was developed, as Appendix B shows.

**Table 10.** Estimation results of OP model.

| Question | ordered Probit Model 1 | | | ordered Probit Model 2 | | |
|---|---|---|---|---|---|---|
| | Coef. | z | p>丨z丨 | Coef. | z | p>丨z丨 |
| 1 | 0.124 | 0.987 | 0.310 | | | |
| 2 | 0.234 | 2.446 | **0.008** | 0.613 | 5.220 | 0.000 |
| 3 | 0.102 | 1.463 | 0.235 | | | |
| 4 | 0.545 | 4.610 | **0.000** | 1.070 | 7.140 | 0.000 |
| 5 | 0.197 | 2.050 | 0.041 | | | |
| 6 | 0.310 | 3.380 | **0.001** | 0.489 | 5.210 | 0.000 |
| 7 | 0.319 | 3.420 | **0.001** | 0.557 | 6.260 | 0.000 |
| 8 | 0.192 | 1.400 | 0.162 | | | |
| 9 | 0.314 | 3.130 | **0.002** | 0.490 | 4.650 | 0.000 |
| 10 | 0.302 | 2.340 | 0.019 | | | |
| 11 | 0.415 | 2.110 | 0.035 | | | |
| 12 | 0.523 | 2.450 | 0.014 | | | |
| 13 | 0.259 | 2.910 | **0.004** | 0.681 | 6.300 | 0.000 |
| 14 | 0.461 | 4.370 | **0.000** | 0.541 | 5.290 | 0.000 |
| 15 | 0.268 | 2.860 | **0.004** | 0.700 | 6.740 | 0.000 |
| 16 | 0.052 | 0.800 | 0.425 | | | |
| 17 | 0-.060 | -0.740 | 0.462 | | | |
| 18 | 0.065 | 0.880 | 0.377 | | | |
| 19 | 0.294 | 2.860 | **0.004** | 0.558 | 4.680 | 0.000 |
| 20 | 0.080 | 1.050 | 0.293 | | | |
| 21 | 0.089 | 0.670 | 0.505 | | | |
| 22 | 0.456 | 3.930 | **0.000** | 0.407 | 3.560 | 0.000 |
| 23 | 0.040 | 0.310 | 0.753 | | | |
| 24 | 0.168 | 1.600 | 0.109 | | | |
| 25 | 0.224 | 2.500 | 0.013 | | | |
| 26 | 0.133 | 1.350 | 0.178 | | | |
| 27 | 0.193 | 2.070 | 0.038 | | | |
| 28 | 0.461 | 3.550 | **0.000** | 0.717 | 5.440 | 0.000 |
| 29 | 0.091 | 0.740 | 0.461 | | | |
| Number of obs | 450 | | | 450 | | |
| Log likelihood | -123.00664 | | | -107.84636 | | |
| Pseudo $R^2$ | 0.7582 | | | 0.7242 | | |
| LR chi2() | LR chi2(29) =163.22 | | | LR chi2(11) =158.95 | | |
| Prob>chi2 | 0.0000 | | | 0.0000 | | |

## 5. Conclusions

This research consisted of two parts. The first part aimed to analyze WLSQ and its influencing factors in the Yangtze Delta from a customer perspective. This analysis was conducted through a questionnaire survey and the use of the SEM method. A total of 750 valid questionnaires were collected, and factor analysis was employed to categorize thirty-two variables into six latent constructs. The results revealed that vessel lockage efficiency had the most significant influence on calculating service quality, followed by waterway lock supporting facilities. On the other hand, lock comfortable conditions and lockage regulations were found to have less importance in influencing customers' perceptions. Additionally, a simplified version of the questionnaire for customers was proposed, utilizing the ordered Probit method to facilitate an ongoing survey.

The second part of the research aimed to analyze service suppliers' perspectives regarding their concerns about waterway lock service performance. This analysis was conducted using the AHP method based on the results obtained from a questionnaire specifically designed for service suppliers. The findings indicated that safety and security were the top priorities for service suppliers, followed by staff skills execution. Therefore, when developing policies, it is crucial to consider the conflicting viewpoints of both customers and service suppliers in order to optimize the lock service.

Furthermore, an interesting finding emerged from the study. Despite assigning relatively high ratings to specific observed variables, a significant number of respondents expressed low overall satisfaction with WLSQ. This discrepancy can be attributed to customers' tendency to prioritize certain observed variables, such as delay, supplement, and service attitude, over others when assessing overall satisfaction. As a result, their evaluations may overlook the impact of other factors, highlighting the limitations of relying solely on a single perspective for evaluation. Moreover, the results indicated that customer satisfaction is influenced by the performance of lock management staff in terms of their attitude and efficiency in problem-solving. Customers do not attach significant importance to the environment or intelligence level of waterway locks. Their primary concern lies in the accuracy and timely release of information for crossing locks, and they perceive the current information release method and operation mode of the locks to be satisfactory. Additionally, a noteworthy finding is that enhancing supporting facilities, staff skills, and safe conditions can significantly improve customers' tolerance for lockage delays.

In the future, research could investigate how environmental changes and technological advancements affect the quality of waterway lock services in the Yangtze Delta. Introducing real-time feedback mechanisms and studying the enduring impacts of infrastructure investments could improve service responsiveness and sustainability.

## CRediT authorship contribution statement

**Wenzhang Yang**: Methodology; Investigation; Funding acquisition; Supervision; Formal analysis; Software; Writing - original draft; Writing - review & editing. **Peng Liao**: Conceptualization; Methodology; Investigation; Funding acquisition; Supervision; Formal analysis; Writing - original draft; Writing - review & editing. **Shangkun Jiang**: Investigation; Formal analysis; Software; Writing - original draft; Writing - review & editing. **Hao Wang**: Supervision; Writing - review & editing.

## Declaration of competing interest

The authors declare that they have no known competing financial interests or personal relationships that could have appeared to influence the work reported in this paper.

**Acknowledgements**

This work was sponsored by the National Natural Science Foundation of China (No. 52172303), the SEU Innovation Capability Enhancement Plan for Doctoral Students (CXJH_SEU 24178), and the Postgraduate Research & Practice Innovation Program of Jiangsu Province (KYCX24_0451).

**Appendix A**

**Table A**

The variable's importance survey questionnaire.

| Criteria Layer | FACTOR | 9 | 7 | 5 | 3 | 1 | 3 | 5 | 7 | 9 | FACTOR | Criteria Layer |
|---|---|---|---|---|---|---|---|---|---|---|---|---|
| NA | WLOE | | | | | | | | | | WLFP | NA |
| | WLOE | | | | | | | | | | WLMS | |
| | WLFP | | | | | | | | | | WLMS | |
| WLOE | Safe& security | | | | | | | | | | Time& convince | WLOE |
| WLFP | Lockage regulation | | | | | | | | | | Supporting facilities | WLFP |
| WLMS | Comfortable conditions | | | | | | | | | | Staff skills | WLMS |

**Note:** NA means these factors belong to criteria layer; WLOE = Waterway lock operation efficiency; WLFP = Waterway lock function performance; WLMS = Waterway lock management specification.

**Appendix B**

**Table B**

The simplified questionnaire.

| Construct | Question | Description | Abbreviation |
|---|---|---|---|
| Sate & security | 1 | Numbers of No Service due to the accidents. | acc_aff |
| Time & convince | 2 | Satisfaction with the vessel's waiting time. | wait_time |
| | 3 | Is the waiting time short than before? | waitime_change |
| | 4 | Satisfaction with the lock's working duration per day. | lock_wktime |
| | 5 | Satisfaction with dealing with the congestion. | cong_deal |
| Lockage regulation | 6 | Satisfaction with the automatic and Lockage regulation. | auto_intlgnt |
| Supporting facilities | 7 | Convenient for waste disposal? | waste_dispo |
| | 8 | Convenient for sewage disposal? | sewage_dispo |
| | 9 | Convenient for adding oil and water? | add_cnvnt |
| Comfortable conditions | 10 | Satisfaction with the ecological environment in lock. | eco_env |
| Staff skills | 11 | Satisfaction with staff's skill of complaint-handling. | cmpln_handle |